\documentclass[conference]{IEEEtran}
\usepackage{cite}
\usepackage{amsmath,amssymb,amsfonts}
\usepackage{algorithmic}
\usepackage{graphicx}
\usepackage{textcomp}
\usepackage{xcolor}
\usepackage{wrapfig}
\usepackage{comment}
\usepackage{amsthm}
\usepackage{longtable}
\usepackage{ upgreek }
\usepackage{amsmath}
\usepackage{graphicx} 
\usepackage{ gensymb }
\usepackage{ dsfont }
\usepackage{tabularx,booktabs}
\usepackage[acronym,nomain,nonumberlist]{glossaries}
\usepackage{glossary-tree}
\usepackage{tabularx}
\newcolumntype{L}{>{\raggedright\arraybackslash}X}

\usepackage{cleveref}
\usepackage{subcaption}
\usepackage{fancyhdr}

\usepackage[ruled,norelsize]{algorithm2e}
\makeatletter
\newcommand{\removelatexerror}{\let\@latex@error\@gobble}
\makeatother
\SetKw{And}{\hspace{\algoskipindent}\itshape and\;}
\SetKwBlock{Condition}{}{}
\SetKw{Or}{\hspace{\algoskipindent}\itshape or\;}

\makeglossaries
\newacronym{5g}{5G}{Fifth Generation}
\newacronym{mmtc}{mMTC}{massive Machine-Type Com\-mu\-ni\-ca\-tions}
\newacronym{urllc}{URLLC}{Ultra-Reliable Low Latency Communications}
\newacronym{embb}{eMBB}{enhanced Mobile Broadband}
\newacronym{tn}{TN}{Terrestrial Network}
\newacronym{dynasat}{DYNASAT}{"Dynamic spectrum sharing and bandwidth-efficient techniques for high-through\-put MIMO Satellite systems"}
\newacronym{ntn}{NTN}{Non-Terrestrial Network}
\newacronym{mc}{MC}{Multi-Connectivity}
\newacronym{iot}{IoT}{Internet of Things}
\newacronym{vr}{VR}{Virtual Reality}
\newacronym{ngso}{NGSO}{Non-Geostationary Orbit}
\newacronym{dsa}{DSA}{Dynamic Spectrum Allocation}
\newacronym{poc}{PoC}{Proof-of-Concept}
\newacronym{3gpp}{3GPP}{3rd Generation Partnership Project}
\newacronym{ue}{UE}{User Equipment}
\newacronym{dc}{DC}{Dual Connectivity}
\newacronym{mn}{MN}{Master Node}
\newacronym{sn}{SN}{Secondary Node}
\newacronym{mrdc}{MR-DC}{Multi Radio-Dual Connectivity}
\newacronym{eutra}{E-UTRA}{Evolved Universal Terrestrial Access}
\newacronym{enb}{eNB}{Evolved Node B}
\newacronym{epc}{EPC}{Evolved Packet Core}
\newacronym{5gc}{5GC}{5G Core}
\newacronym{nr}{NR}{New Radio}
\newacronym{engnb}{en-gNB}{en-Next Generation Node B}
\newacronym{ngenb}{ng-eNB}{Next Generation eNB}
\newacronym{ca}{CA}{Carrier Aggregation}
\newacronym{cc}{CC}{Component Carrier}
\newacronym{rrm}{RRM}{Radio Resource Management}
\newacronym{pdcp}{PDCP}{Packet Data Convergence Protocol}
\newacronym{mac}{MAC}{Media Access Control}
\newacronym{rrc}{RRC}{Radio Resource Control}
\newacronym{cn}{CN}{Core Network}
\newacronym{mcg}{MCG}{Master Cell Group}
\newacronym{scg}{SCG}{Secondary Cell Group}
\newacronym{rf}{RF}{Radio Frequency}
\newacronym{ngran}{NG-RAN}{Next Generation Radio Access Network}
\newacronym{gnbcu}{gNB-CU}{gNB-Centralized Unit}
\newacronym{gnbdu}{gNB-DU}{gNB-Distributed Unit}
\newacronym{leo}{LEO}{Low Earth Orbit}
\newacronym{geo}{GEO}{Geostationary Orbit}
\newacronym{ran}{RAN}{Radio Access Network}
\newacronym{hetnet}{HetNet}{Heterogeneous Networks}
\newacronym{rsrp}{RSRP}{Reference Signal Received Power}
\newacronym{ns3}{ns-3}{Network Simulator 3}
\newacronym{gnb}{gNB}{Next Generation Node B}
\newacronym{e2e}{E2E}{End-to-End}
\newacronym{pgw}{PGW}{Packet Network Data Gateway}
\newacronym{sgw}{SGW}{Serving Gateway}
\newacronym{amf}{AMF}{Access and Mobility Management Function}
\newacronym{upf}{UPF}{User Plane Function}
\newacronym{ra}{RA}{Random Access}
\newacronym{4g}{4G}{Fourth Generation}
\newacronym{ap}{AP}{Access Point}
\newacronym{srs}{SRS}{Sounding Reference Signal}
\newacronym{udp}{UDP}{User Datagram Protocol}
\newacronym{sls}{SLS}{System Level Simulator}
\newacronym{kpi}{KPI}{Key Performance Indicator}
\newacronym{ecdf}{eCDF}{empirical Cumulative Distribution Function}
\newacronym{tcp}{TCP}{Transmission Control Protocol}
\newacronym{ahp}{AHP}{Analytic Hierarchy Process}
\newacronym{rat}{RAT}{Radio Access Technology}
\newacronym{sinr}{SINR}{Signal-to-Interference-plus-Noise Ratio}
\newacronym{ts}{TS}{Technical Specification}
\newacronym{tr}{TR}{Technical Report}
\newacronym{lan}{LAN}{Local Area Network}
\newacronym{rnti}{RNTI}{Radio Network Temporary Identifier}
\newacronym{sib}{SIB}{System Information Block}
\newacronym{mib}{MIB}{Master Information Block}
\newacronym{nlos}{NLOS}{Non-Line of Sight}
\newacronym{rng}{RNG}{Random Number Generator}
\newacronym{sue}{SUE}{Spectral Utilization Efficiency}
\newacronym{rb}{RB}{Resource Block}
\newacronym{re}{RE}{Resource Element}
\newacronym{ewma}{EWMA}{Exponential Weighted Moving Average}
\newacronym{sdap}{SDAP}{Service Data Adaption Protocol}
\newacronym{ho}{HO}{Handover}
\newacronym{pdu}{PDU}{Protocol Data Unit}
\newacronym{wa}{WA}{Wraparound}
\newacronym{cbr}{CBR}{Constant Bit Rate}

\def\BibTeX{{\rm B\kern-.05em{\sc i\kern-.025em b}\kern-.08em
    T\kern-.1667em\lower.7ex\hbox{E}\kern-.125emX}}

\renewcommand\IEEEkeywordsname{Keywords}

\makeatletter
\newcommand{\linebreakand}{%
  \end{@IEEEauthorhalign}
  \hfill\mbox{}\par
  \mbox{}\hfill\begin{@IEEEauthorhalign}
}
\makeatother

\fancypagestyle{FirstPage}{

\lfoot{\copyright 2023 IEEE. Personal use of this material is permitted. Permission from IEEE must be obtained for all other uses, in any current or future media, including reprinting/republishing this material for advertising or promotional purposes, creating new collective works, for resale or redistribution to servers or lists, or reuse of any copyrighted component of this work in other works. DOI 10.1109/WoWMoM57956.2023.00073}

}

\begin{document}
\bstctlcite{IEEEexample:BSTcontrol}
\newtheorem{thm}{Theorem} 
\theoremstyle{definition}
\newtheorem{remark}[thm]{Remark}
\newtheorem{defn}[thm]{Definition}
\theoremstyle{plain}
\newtheorem{thr}[thm]{Theorem}
\newtheorem{prop}[thm]{Proposition}
\newtheorem{kor}[thm]{Corollary}

\title{Satellite-Assisted Multi-Connectivity in Beyond 5G}

\author{
\IEEEauthorblockN{Mikko Majamaa\IEEEauthorrefmark{1}\IEEEauthorrefmark{2}, Henrik Martikainen\IEEEauthorrefmark{1}, Jani Puttonen\IEEEauthorrefmark{1} and Timo Hämäläinen\IEEEauthorrefmark{2}}

\IEEEauthorblockA{
\IEEEauthorrefmark{1}\textit{Magister Solutions, Jyv\"{a}skyl\"{a}, Finland} \\
email: \{firstname.lastname\}@magister.fi
}

\IEEEauthorblockA{
\IEEEauthorrefmark{2}\textit{Faculty of Information Technology, University of Jyv\"{a}skyl\"{a}, Finland} \\
email: timo.t.hamalainen@jyu.fi}

}

\maketitle

\begin{abstract}

Due to the ongoing standardization and deployment activities, satellite networks will be supplementing the 5G and beyond Terrestrial Networks (TNs). For the satellite communications involved to be as efficient as possible, techniques to achieve that should be used. Multi-Connectivity (MC), in which a user can be connected to multiple Next Generation Node Bs simultaneously, is one such technique. However, the technique is not well-researched in the satellite environment. In this paper, an algorithm to activate MC for users in the weakest radio conditions is introduced. The algorithm operates dynamically, considering deactivation of MC to prioritize users in weaker conditions when necessary. The algorithm is evaluated with a packet-level 5G non-terrestrial network system simulator in a scenario that consists of a TN and transparent payload low earth orbit satellite. The algorithm outperforms the benchmark algorithms. The usage of MC with the algorithm increases the mean throughput of the users by 20.3\% and the 5th percentile throughput by 83.5\% compared to when MC is turned off.

\end{abstract}

\begin{IEEEkeywords}
Non-Terrestrial Networks (NTNs), 6G, Low Earth Orbit (LEO) satellite, throughput enhancement, satellite network simulator
\end{IEEEkeywords}

\section{Introduction}
\label{sec:introduction}

\thispagestyle{FirstPage}

5G and beyond Non-Terrestrial Networks (NTNs) are emerging to complement the Terrestrial Networks (TNs). NTNs can help to provide ubiquitous 5G and beyond services to unserved and underserved areas and can provide load balancing to overloaded TNs. Further, new disruptive use cases utilizing satellite communications are emerging such as direct-to-handheld satellite access.

History was made in 3rd Generation Partnership Project (3GPP) Release 17, finalized in March 2022. The release included NTN communications in the 3GPP mobile communication standards for the first time. NTN standardization in 3GPP started in Release 15 and 16 respectively with reports concerning study \cite{38811} and solutions \cite{38821} for New Radio (NR), the air interface of 5G, to support NTNs. These reports identified issues inherent in NTNs, e.g., high propagation delays, interference, regulatory issues, and high Doppler shifts due to the high velocity of satellites. Release 17 included a set of basic features for NR to support Low Earth Orbit (LEO) and Geostationary Earth Orbit (GEO) satellite communications with implicit support to High Altitude Platform Station (HAPS) and Air-to-Ground (A2G) scenarios. Release 18 marks the beginning of the 5G-Advanced (5G-A) era and the standardization toward 6G. NTN-wise, Release 18 will enhance the NR operations to support NTNs, e.g., by improving coverage for handheld terminals, studying deployment in higher frequencies (above 10 GHz), and enhancing mobility and service continuity aspects.

For the emerging satellite communications to be as efficient as possible, techniques to achieve that should be used. Multi-Connectivity (MC) in which a user can be connected to more than one base station is one such technique. MC can be used, e.g., in load balancing, peak throughput enhancement, providing throughput enhancement to users in weak radio conditions, and in aid in soft Handovers (HOs). For TNs, MC is specified by 3GPP in \cite{37340}. MC in NTNs remains to be specified but is a Release 19 candidate. For that, there is a need for research related to MC in NTNs.

Next, related work is discussed. NTN standardization journey in 3GPP is elaborated on in \cite{ntnsurvey}. It is concluded that the relevance of NTNs in 5G and beyond networks is expected to increase. A comprehensive survey of MC in mobile networks is provided in \cite{s22197591}. Enabling technologies, existing standards, and current solutions are reviewed. It is concluded that MC is a promising solution for the ever-increasing bandwidth, reliability, and latency requirements.

Only a few works consider MC in NTNs. MC  in NTNs is briefly discussed in \cite{38821}. It is stated that the nodes involved can be NTN-based Next Generation Node Bs (gNBs) and TN-based nodes providing Evolved Universal Terrestrial Radio Access (E-UTRA) (4G)/NR access. The NTN-based gNBs are considered to have either transparent or regenerative payloads. Satellites with transparent payloads act as analog radio frequency repeaters whereas satellites with regenerative payload (part of) gNB is onboard the satellite. The authors have previously researched MC in NTNs in \cite{9911400, 9941510, majamaa2022multiconnectivity}, where MC activation and traffic steering algorithms for throughput enhancement were introduced. The algorithms were evaluated in a scenario with transparent payload LEO satellites. MC between TN and NTN is considered in \cite{9621102}. However, Non-Geostationary Orbit (NGSO) satellites are not considered, nor direct to handheld scenarios. MC for reliability in TN-NTN scenario with a special type of receiver antenna is researched in \cite{10008752}. This leaves space for research, e.g., of MC between satellites, in a scenario with direct-to-handheld, and MC to improve throughput. An uplink scheduling strategy for MC with multi-orbit NTN is introduced in \cite{10008521}, leaving room for further research on downlink scheduling.

Although some research regarding MC in NTNs has been conducted, there is a need to enhance the MC-related algorithms. Further, direct-to-handheld satellite access, especially in the NGSO case, in combination with MC needs research. In this paper, we introduce an MC activation algorithm that aims to enhance the throughputs of the weakest users based on their channel conditions. The algorithm adapts dynamically, taking into account load-based deactivation of MC to prioritize users in weaker radio conditions. The algorithm is evaluated by packet-level system simulations in a scenario with a TN, transparent payload LEO satellite, and users with handheld devices.

The rest of the paper is organized as follows. In the next section, MC in NTNs is discussed. In Section~\ref{sec:mcalgorithms}, the developed MC activation and used traffic split algorithms are introduced. Section~\ref{sec:simulations} describes the simulations to evaluate MC in a scenario that consists of a satellite and TN. Finally, the paper is concluded in Section~\ref{sec:conclusions}.

\section{Multi-Connectivity in Non-Terrestrial Networks}
\label{sec:mcinntns}

Multi-Radio Dual-Connectivity (MR-DC) \cite{37340} is a generalization of E-UTRA DC \cite{36300}. In MR-DC, one node provides NR access and one either NR or E-UTRA access. NR-DC is a form of MR-DC in which both nodes are providing NR access, i.e., both nodes are gNBs. The MC considered in this work is NR-DC in which one gNB acts as a Master Node (MN) and the other as a Secondary Node (SN). In principle, there could be more than one SNs. However, limitations could be posed by the hardware and software requirements of the User Equipment (UE).

In NR-DC, the MN and SN are connected for Control Plane (CP) signaling through the Xn-C interface. Each of the gNBs has its own Radio Resource Control (RRC) state, whereas the UE has a single RRC state and a single CP connection toward the Core Network (CN). SRB3 interface can be configured between the SN and UE for RRC signaling, or RRC Protocol Data Units (PDUs) from the SN to the UE can be forwarded through the MN. The connection between the MN and SN for User Plane (UP) data exchange is handled through the Xn-U interface. When data for the UE arrives at the MN’s Packet Data Convergence Protocol (PDCP) layer, the MN can either forward it to its lower layers for transmission or forward the data through the Xn-U interface to the SN which then sends it to the UE. From the UE perspective, three types of bearers exist – Master Cell Group (MCG), Secondary Cell Group (SCG), and split bearers. In these bearers, resources of the MN, SN, or both are involved, respectively.

The SN addition process is initiated by the UE’s current serving gNB. Based on some trigger (e.g., signal strength or distance), the UE’s current serving gNB sends an SN addition request to a candidate SN. The candidate can reject or acknowledge the request. If the request is acknowledged, the MN (i.e., the UE’s current serving gNB) informs the UE by RRC reconfiguration required message. The UE does the needed reconfigurations, responds to the MN, and the MN informs the SN that the reconfiguration is complete. After the SN addition process, the MN can start to forward data to the SN which then sends it to the UE.

\begin{figure}[htb!]
    \centering
    \includegraphics[width=\linewidth]{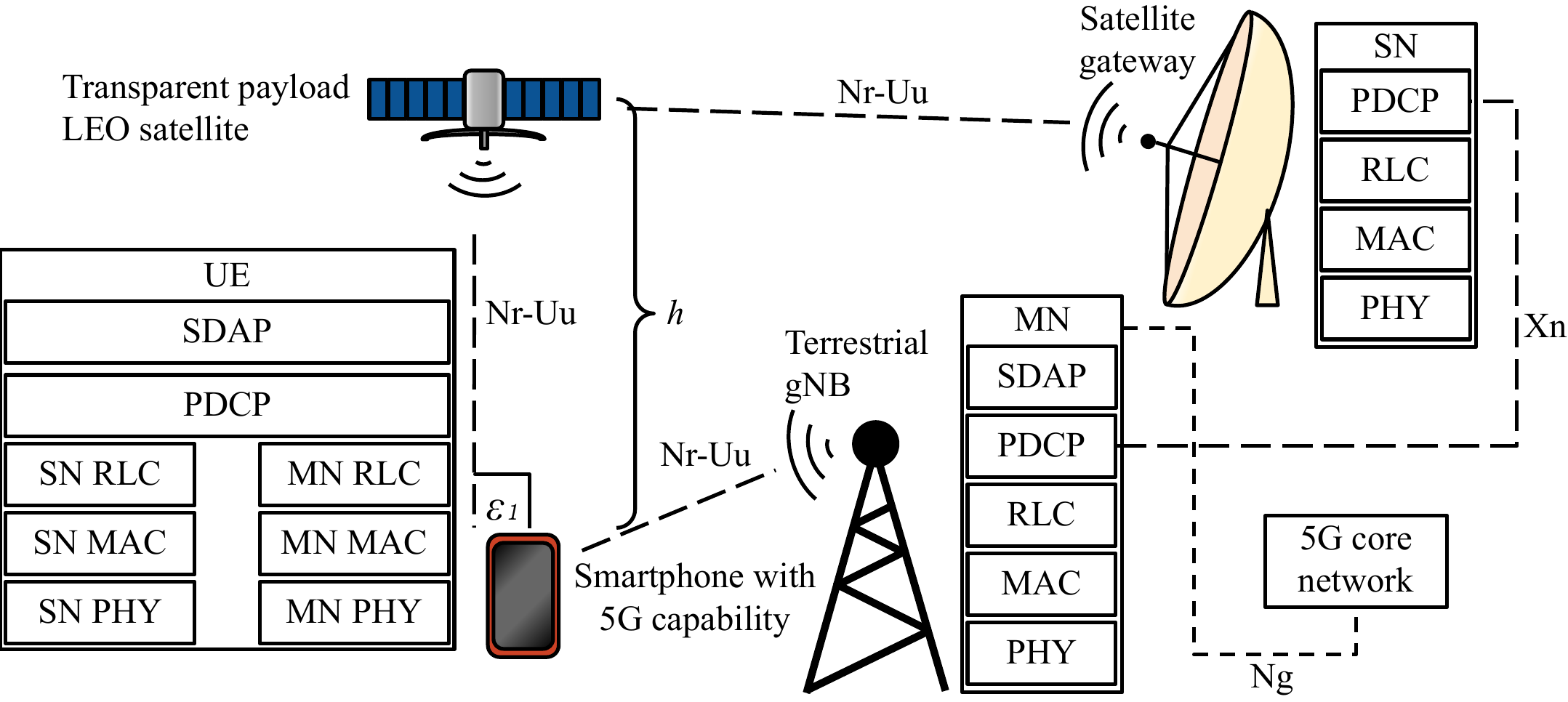}
    \caption{MC involving a terrestrial gNB and NTN gNB.}
    \label{fig:tnntnmcillustrated}
\end{figure}

MC involving a terrestrial gNB and NTN gNB is illustrated in Fig.~\ref{fig:tnntnmcillustrated}. In the figure, the UE is a smartphone with 5G capability. The UE is connected to a terrestrial gNB and NTN gNB involving a transparent payload LEO satellite. Since the payload of the satellite is transparent,  both the service and feeder links implement 5G Nr-Uu radio interface. The gNBs are connected through the Xn interface for CP and UP signaling. The figure also shows the 5G UP protocol stacks for the UE and gNBs. The UE must be able to receive transmissions from both the gNBs. The transmissions are then aggregated at the PDCP layer. The elevation angle ($\epsilon_1$) changes as the satellite moves and the altitude ($h$) depends on the satellite's orbit. 

The gNBs can reside close to each other or even on different continents. However, in the latter case, there might be a significant difference in the propagation delays. Different propagation delays are likely to lead to out-of-order packet reception at the UE's PDCP layer. The PDCP has reordering and buffering capabilities which may be exceeded with a large propagation delay difference. This should be considered either at the UE or gNB side such that no buffer overflows occur. Operators may also disable MC between gNBs too far from each other.

{}

\section{Multi-Connectivity Algorithms}
\label{sec:mcalgorithms}
\subsection{Secondary Node Addition}
\label{sec:snaddition}

MC includes the process of SN addition and after that traffic splitting between the MN and SN. In this work, MC activation for a UE is based on its Modulation and Coding Scheme (MCS) related to the transmission of the current serving gNB. MCS defines the number of useful bits that can be transmitted in a Resource Element (RE). It is selected by the gNB and signaled to the UE in Downlink Control Information (DCI). MCS depends on radio conditions. The gNB transmits Channel State Information-Reference Signal (CSI-RS) and the UE responds with CSI feedback from which the used MCS can be deduced. In 5G NR, there are 32 MCS indices (0-31), and higher MCS index generally indicates better radio conditions.

The goal here is to enhance the throughputs of the users with the weakest conditions. The weak users are identified by a low MCS. Another way to identify weak users could be, e.g., by the distance to the serving cell. However, because of, e.g., shadowing conditions, the distance may not be a sufficient condition. Further rationale for the use of MCS as a criterion for SN addition is that it is already available from the CSI signaling. Note that MCS may be an unstable metric when Fast Fading (FF) is high, i.e., MCS may vary from measurement to measurement. In such scenarios, averaging MCS over a period should be considered.

Algorithm 1 describes the MC activation process. The algorithm is run locally on every gNB that has UEs connected to it that may require an SN. Whether a request for SN addition for a UE is sent is evaluated periodically by a controller in each gNB. On every such evaluation step, the UEs that do not have secondary connections are iterated in ascending order based on the MCS that is used by the gNB in the transmission towards each of the UEs. During the iteration, if a UE has a higher MCS than a predefined maximum MCS value for MC activation, the SN addition evaluation is terminated for this evaluation step because the succeeding UEs can neither have the required MCSs. If the MCS threshold is met, a candidate gNB for SN addition is searched.

Reference Signal Received Power (RSRP) is the average power of the resource elements that carry secondary synchronization signals. RSRP measurements are measured by the UE and signaled to the serving gNB. These measurements give insight into signal strengths of possible HO/SN addition targets. In the algorithm, finding a candidate for SN addition corresponds to finding the highest RSRP measurement related to the gNBs for which there exist non-outdated measurement reports and to which no SN addition requests have been sent during the last allowed request-sending period ($t_\textnormal{req\_period}$). If such a candidate is found, an SN addition request is sent if the RSRP threshold (RSRP$_\textnormal{th}$) for SN addition is met. 

When the SN addition request is received by the candidate, it is rejected if another SN addition request was accepted, from any gNB, during the last request-acknowledging period ($t_\textnormal{add}$). Otherwise, the gNB checks its load status and if free capacity exists, the request is acknowledged. If no capacity exists, the gNB searches for a UE, which it is serving as an SN, with the highest MCS related to the UE’s MN transmission. If a secondary connection UE is found and the MCS is higher than the UE’s whose SN addition is considered, the secondary connection is released. In consequence, the SN addition request for the considered UE is acknowledged due to the freed resources. The info about the MCS related to the MN transmission can be piggybacked in the SN addition request and stored by the SN. If the MCS changes, the MN can signal the new MCS through an Xn message. At the end of each evaluation step, a small random delay on when to perform the next evaluation is introduced to ensure that the requests from different gNBs do not always arrive in the same order to guarantee fairness.

\begin{figure}[!t]
 \removelatexerror
  \begin{algorithm}[H]
   \label{algorithm:snaddition}
   \caption{Secondary Node Addition Algorithm}
   \While{$t_{\textnormal{eval}} < t_\textnormal{sim}$}
    {
    $bestRsrp = -\textnormal{INF}$ \;
    $bestCellId = 0$ \;
    
    Order the single-connectivity users in ascending order based on MCS to obtain set $\mathfrak{U}$ \;  
     
    \ForEach{$ue \in \mathfrak{U}$}
    {
        \uIf{$ue.\textnormal{MCS} > \textnormal{MCS$_\textnormal{th}$}$}
        {
            break \;
        }
        \ForEach{$m \in ue.\textnormal{neighborCellMeasurements}$}
        {
            $i = m.\textnormal{cellId}$ \;
            \DontPrintSemicolon
            \uIf{$m.\textnormal{RSRP} > bestRsrp$ \textnormal{\textbf{and}} $t - t_{\textnormal{last},i} \ge t_\textnormal{req\_period}$ \textnormal{\textbf{and}} $t - m.\textnormal{time} \leq t_\textnormal{val}$}
           {
                \PrintSemicolon
                $bestRsrp = m.\textnormal{RSRP}$ \;
                $bestCellId = i$ \;
           }
        }
        \uIf{$bestCellId > 0$ \textnormal{\textbf{and}} $m.\textnormal{RSRP} \ge \textnormal{RSRP$_\textnormal{th}$}$}
        {
            Send SN addition request to $bestCellId$ \;
            /* At the candidate node $bestCellId$: */ \\
            \uIf{$t - t_{\textnormal{last}} \leq t_\textnormal{add}$}
             {
                Reject the SN addition \;
             }
            \uIf{$L \leq L_{th}$}
             {
                Acknowledge the SN addition \;
             }
            \uElse
            {
                Find UE $k$ with the highest MCS (to its MN) that the gNB is serving as an SN. If the MCS is higher than the UE's MCS whose SN addition is considered, release the secondary connection of $k$, and acknowledge the considered SN addition \;
            }
            /* At the candidate node $bestCellId$ \textasciicircum \textasciicircum \ */ \\
        }
    }
    $t_\textnormal{del}=\textnormal{RAND} \cdot 1 \textnormal{ms}$ \;
    $t_{\textnormal{eval}} = t_{\textnormal{eval}} + t_{\textnormal{eval\_period}} - t_\textnormal{prev} + t_\textnormal{del}$ \;
    $t_\textnormal{prev} = t_\textnormal{del}$ \;
    }

  \end{algorithm}
\end{figure}

\subsection{Traffic Splitting}

After an SN has been added to a user, the decision of how to split the traffic between the MN and SN must be made. Here, traffic splitting is based on data requests from the SN to MN. The algorithm was introduced in \cite{majamaa2022multiconnectivity} and works as follows. The SN sends periodic data requests to the MN on per user basis through the Xn interface. Based on the load left over from the primary connection users, the SN computes the theoretical amount of data that it can forward to a secondary connection user using the Shannon’s formula. The amount of data that an SN $i$ requests from an MN to a user $j$ to send during a period is computed as

\begin{equation}
\label{eq:amountofdata}
D=\alpha\cdot\frac{1-L_{\textnormal{pr},i}}{n_{\textnormal{s},i}}\cdot B \cdot \log_2 (1+\textnormal{SINR}_{ij}) \cdot (\Delta t + t_{\textnormal{off}}),
\end{equation}

where $\alpha$ accounts for the implementation losses (defined as 0.6 in \cite{38803}), $L_{\textnormal{pr},i}$ is the load posed by the primary connection users of node \textit{i}, $n_{\textnormal{s},i}$ is the number of secondary connection users at node $i$, $B$ is the total available bandwidth of the SN, SINR$_{ij}$ is the Signal-to-Interference-and-Noise-Ratio (SINR) of user $j$ related to node $i$, $\Delta t$ is the data request interval, and $t_{\textnormal{off}}$ is the data request period offset to request data ahead of time, e.g., to account for lost requests.

The MN stores the data requests and when data to a user arrives at the MN’s PDCP layer, the MN checks whether it has a valid request for the user. If it has, it  forwards the data through the Xn interface to the SN. If it doesn't, the MN forwards the data to the lower layers of its protocol stack to send to the user.

\section{Simulations}
\label{sec:simulations}

\subsection{5G Non-Terrestrial Network Simulator}

The introduced MC implementation is evaluated by system simulations. The used simulator is a 5G NTN System-Level Simulator (SLS) \cite{ntn} that is built on top of Network Simulator 3 (ns-3) \cite{Riley2010} and its 5G LENA module \cite{Patriciello2019AnES}. ns-3 is a discrete-event non-real-time packet-level network simulator mainly used for educational and research purposes. Users may add new modules to the simulator. 5G LENA  is one such module that is used to simulate 5G networks. However, 5G LENA cannot simulate NTNs. In the 5G NTN SLS, 5G LENA was used as a starting point and the necessary components to simulate NTNs were implemented.

5G LENA implements NR Physical (PHY) and Medium Access Control (MAC) features but the upper layers of the UE/gNB stack are reused from the ns-3 Long Term Evolution (LTE) module \cite{lenalte}. The ns-3 core provides the higher layers (e.g., transport and network). The link layer is abstracted with Link-to-System (L2S) mapper and Modulation and Coding (MODCOD)-specific SINR to Block Error Rate (BLER) mapping curves. SINR is computed for each packet and using the mapper, BLER is deduced.

In the framework of past R\&D efforts, the simulator has been calibrated using the system-level calibration scenarios from \cite{38821}. Channel and antenna/beam modeling from \cite{38811} have been implemented in the simulator, as well as global coordinate system and calibration scenarios from \cite{38821}. The calibration scenarios work as a baseline for parameterizations, but they can be adjusted as desired. The scenarios provide different assumptions, e.g., bands (S-band/Ka-band), terminal types (VSAT, handheld), and frequency reuse patterns (reuse 1, 3, 2+2). Further, hybrid TN-NTN scenarios can be studied. MC has been implemented in the simulator following the specifications in \cite{37340}.

\subsection{Scenario and Assumptions}

The considered scenario consists of a TN and transparent payload LEO satellite. The TN gNBs and NTN gNB are considered to be close to each other (thus, the delay in the Xn interface is considered negligible). The motivation for the selected scenario is that the NTN can provide service continuity but here it can also be used for throughput enhancement. The TN is comprised of three sites each with three sectors. The Inter-Site Distance (ISD) is 7.5 km. The satellite’s center beam is steered in the middle of the TN sites (quasi-earth fixed beam deployment is considered). Only one actual NTN beam is considered because it can cover the whole TN area with the typical LEO satellite footprint of 100-1000 km \cite{38821}. Further, two tiers of Wraparound (WA) beams are considered with Frequency Reuse Factor 3 (FRF3). The WA beams serve one full buffer user each. WA is used only to introduce more realistic interference; thus, the WA users are left out of statistics collection.

The satellite flies over the TN and offers load balancing and throughput enhancement for the TN gNBs. For comparison, when MC is turned off, only the TN serves the users. When MC is turned on, the NTN node is the SN for the users with MC activated. Note that this need not be the case, i.e., the NTN node could be the MN as well. However, in this work, the focus is on MC in which TN gNBs are the MNs in relation to the NTN gNB. 

\begin{figure}[htb!]
    \centering
    \includegraphics[trim={0 0 0 0},clip, width=\linewidth]{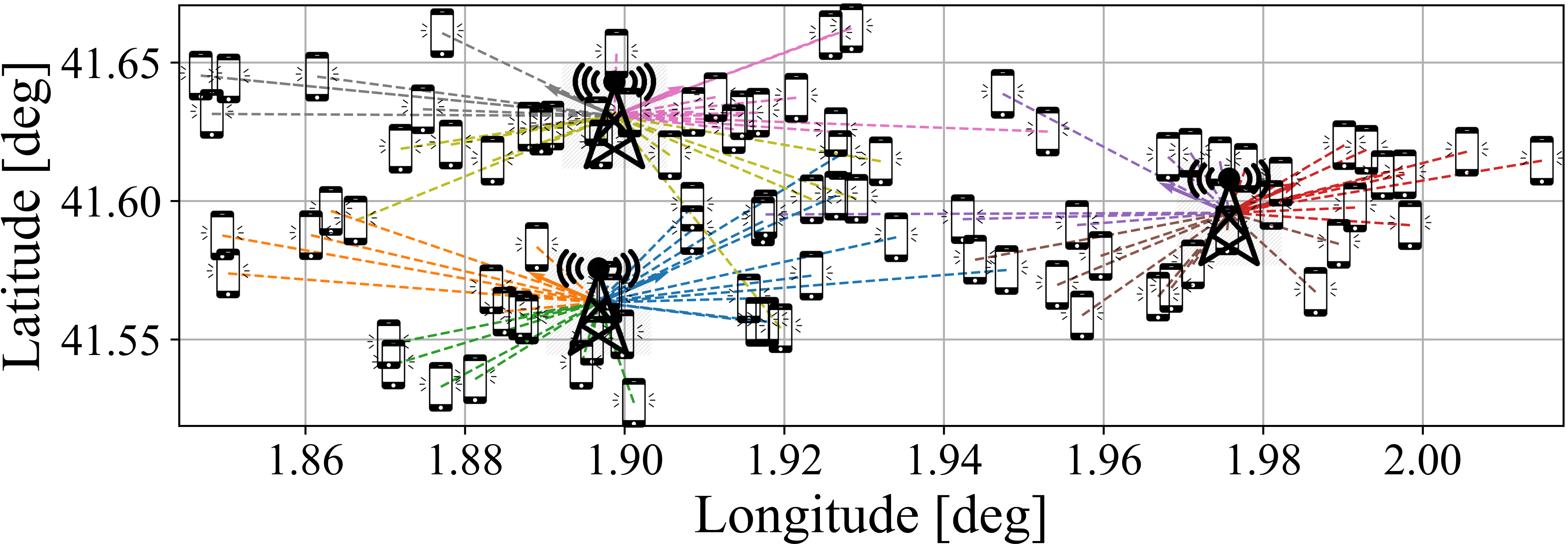}
    \caption{TN considered has three sites with three sectors each. The satellite’s beam is steered at the center of the TN sites.}
    \label{fig:scenario}
\end{figure}

The TN is illustrated in Fig.~\ref{fig:scenario}. Ten users are placed randomly around each TN sector. The dashed lines depict the UE-gNB connections. The satellite’s beams and TN sectors’ bandwidths are 10 MHz each. The simulation time is 5.0 s. The warmup time is 2.5 s. The statistics are collected after the warmup so that the system has time to reach a steadier state. The satellite is considered moving with a velocity of 7.56 km/s at 600 km altitude \cite{38811}. Due to the relatively insignificant movement of the UEs, they are considered stationary. Channel conditions in the TN are dynamic in the sense that the Non-Line of Sight (NLOS) probability is higher for longer distances between a UE and gNB. Because of the high elevation angle in relation to the satellite, the users are considered to have the satellite in LOS. For the same reason, FF is considered negligible to the analysis of MC in this work and is disabled. The users are considered to require Constant Bit Rate (CBR) traffic with User Datagram Protocol (UDP) at 3200 kbps. In this work, only downlink is considered.

{}

The different setups for the simulations are run 15 times each with different \gls{rng} seeds that lead to, e.g., different UE positions. For the distribution statistics, the results are combined, whereas for the scalar statistics they are averaged. The different setups correspond to running the simulations without MC (only the TN serves the users) and with MC using different SN addition algorithms.

The first SN addition algorithm considered is an RSRP-based algorithm for SN addition, which is a common practice, that adds a secondary connection to a user when the RSRP threshold (RSRP$_\textnormal{th}$) towards the candidate SN is met. The second SN addition algorithm used is based on recognizing the users’ need for SN addition by Transmission (Tx) Buffer Occupancy (BO) towards the user at the serving gNB ― i.e., when the Tx buffer is occupied to a certain threshold ($O_\textnormal{th}$) the user is considered to need an SN. Further, the signal related to the candidate SN must be above the RSRP threshold. The algorithm is further detailed in \cite{majamaa2022multiconnectivity} and is referred to as the BO-based SN addition algorithm. The third SN addition algorithm is the one introduced in Section~\ref{sec:snaddition} and is referred to as the MCS-based SN addition algorithm.

For all the SN addition algorithms, the RSRP threshold is -111 dBm. It is chosen low enough so that enough users have the opportunity to have MC activated for them but high enough so that the secondary connection wouldn’t be too weak. The periods that SN additions may be sent to a candidate SN and in which SN additions are acknowledged by a candidate are set to 100 ms. This is to ensure that the recently added secondary connections have posed their effect on the load before acknowledging new ones. For the MCS-based SN addition, the SN addition evaluation step is 10 ms and the MCS threshold is 15 (which is chosen empirically). The most important simulation parameters are presented in Table~\ref{table:params}.

\begin{table}[]
\caption{Simulation parameters.}
\label{table:params}
\begin{tabularx}{\linewidth}{l|L}
\hline
\textbf{Parameter}               & \textbf{Value}                                   \\ \hline
Simulation time ($t_\textnormal{sim}$)                & 5.0 s                                            \\
Warmup time                      & 2.5 s                                            \\ 
Satellite mobility               & Moving                                           \\ 
UE mobility                      & Stationary                                       \\ 
Beam deployment                  & Quasi-Earth Fixed                                \\ 
Satellite starting position      & Lat: 41.59°, Lon: 1.74°               \\ 
NTN channel condition            & LOS                                              \\ 
TN channel condition             & Dynamic LOS                                     \\ 
Number of TN sites               & 3                                                \\ 
Sectors per site                 & 3                                                \\ 
UEs per sector                     & 10                                               \\ 
ISD                              & 7.5 km                                           \\ 
TN deployment                    & Rural                                            \\ 
Bandwidth per NTN beam ($B$)       & 10 Mhz                                           \\ 
Bandwidth per TN sector          & 10 Mhz                                           \\ 
NTN carrier frequency            & 2 GHz (S-band)                                   \\ 
Satellite orbit                  & LEO 600 km                                       \\ 
Satellite parameter set          & Set 1 \cite[Table 6.1.1.1-1]{38821}                   \\ 
UE antenna type                  & Handheld                                         \\ 
Traffic                     & CBR with UDP                                              \\ 
Required rate per UE                  & 3200 kbps                                            \\ 
FF                      & Disabled                                         \\ 
SN addition RSRP threshold (RSRP$_\textnormal{th}$)     & -111 dBm                                         \\ 
SN addition evaluation step ($t_\textnormal{eval}$)    & 10 ms                                            \\ 
SN addition load threshold ($L_\textnormal{th}$)     & 0.975                                     \\ 
SN addition request period ($t_\textnormal{req\_period}$)     & 100 ms                                           \\ 
SN addition acknowledge period ($t_\textnormal{add}$) & 100 ms                                           \\ 
Tx BO threshold ($O_\textnormal{th}$)  & 0.8                                             \\ 
Scheduler                        & Round Robin (primary connection UEs prioritized) \\ 
RSRP measurement report interval & 120 ms                                           \\ 
RSRP measurement validity time ($t_\textnormal{val}$) & 120 ms                                           \\ 
Data request period ($\Delta t$)            & 25 ms                                            \\ 
Data request period offset ($t_\textnormal{off}$)     & 25 ms                                            \\ 
RNG runs                         & 15                                               \\ \hline
\end{tabularx}
\end{table}

\subsection{Results}

\begin{figure}[htb!]
    \centering
    \includegraphics[width=\linewidth]{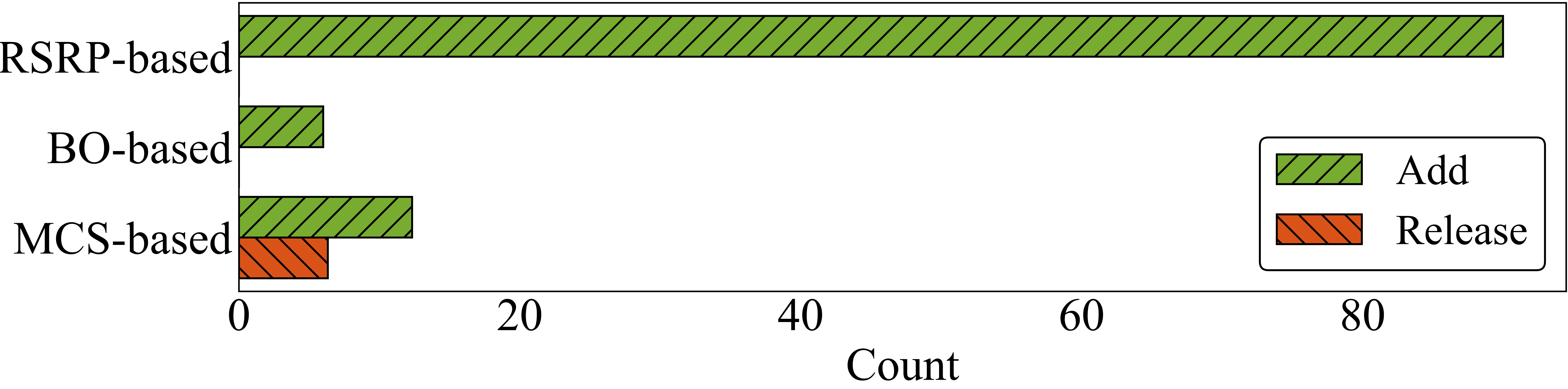}
    \caption{SN adds and releases with the different MC settings.}
    \label{fig:snaddition}
\end{figure}

The SN addition and release counts with the different MC settings are depicted in Fig.~\ref{fig:snaddition}. The MCS-based SN addition algorithm demonstrates an increase in the number of secondary connection additions compared to the BO-based SN addition algorithm. The MCS-based SN addition algorithm adds 12.3 secondary connections on average whereas the BO-based SN addition algorithm adds on average 6 secondary connections. This is because the MCS-based algorithm prioritizes users in worse channel conditions, i.e., it releases UEs in better conditions if needed. The MCS-based SN addition algorithm performs 6.3 releases on average. The number of concurrent secondary connections converge for these two algorithms. The RSRP-based SN addition algorithm adds a secondary connection to all the users since the RSRP threshold is met. This leads to suboptimal SN additions, as shall be seen from the throughput statistics.

{}

\begin{figure}[htb!]
    \centering

    \begin{subfigure}{0.613\linewidth}
        \includegraphics[trim={0 0 0 0},clip, width=\linewidth]{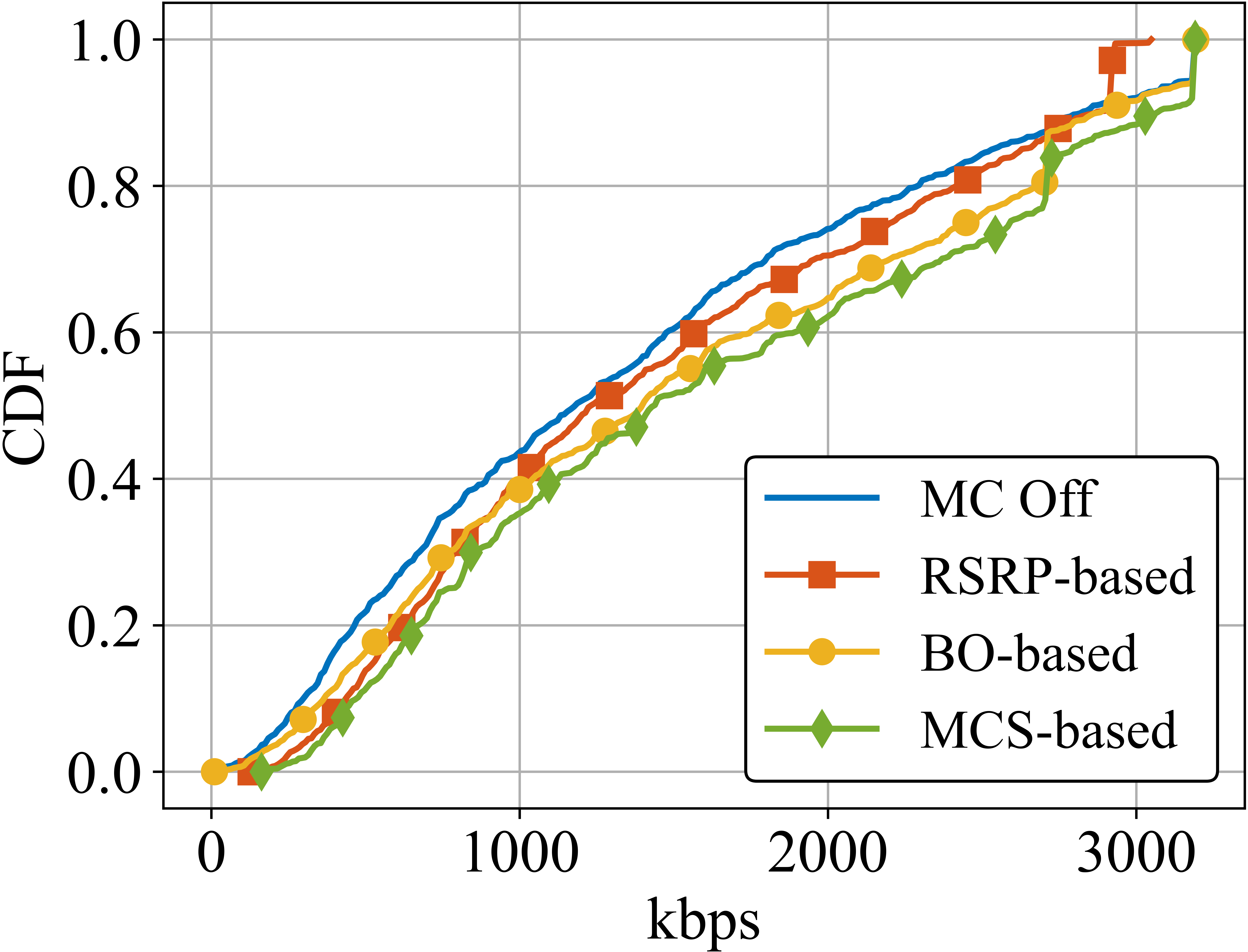}
          \caption{CDF of the throughputs.}
          \label{fig:tp}
      \end{subfigure}
      \hfill
      \begin{subfigure}{0.367\linewidth}
        \includegraphics[trim={0 0 0 0},clip, width=\linewidth]{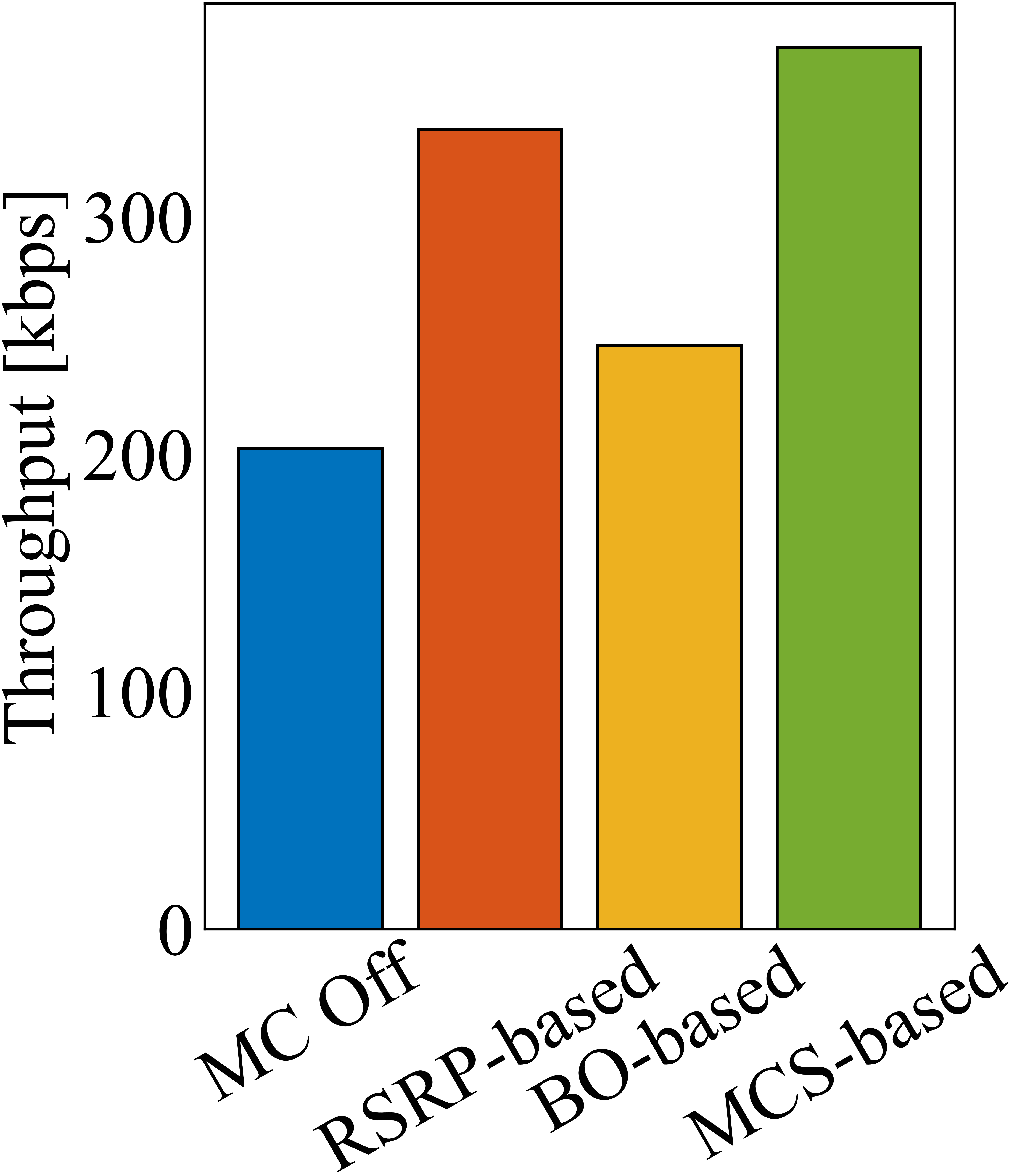}
          \caption{5th percentile throughputs.}
          \label{fig:tp5}
      \end{subfigure}

      \caption{Application throughputs of the users.}
\end{figure}

Fig.~\ref{fig:tp} shows the Cumulative Distribution Function (CDF) of the users' application throughputs in the different simulation setups. It can be observed that turning on MC with any of the algorithms enhances the mean throughput of the users, as expected. The MCS-based SN addition algorithm increases the mean throughput by 20.3\% compared to when MC is turned off. For convenience, the throughput statistics are summarized in Table~\ref{table:tp}. 

The 5th percentile application throughputs of the users in the different simulation setups are captured in Fig.~\ref{fig:tp5}. It can be seen that the RSRP-based SN addition algorithm performs better compared to the BO-based SN addition algorithm in terms of the 5th percentile throughputs. However, this comes with a cost of lower mean throughput and the high overhead of managing the large number of secondary connections. Compared to MC turned off, MC on with RSRP-based addition, and MC on with BO-based addition, the MCS-based SN addition increases the 5th percentile throughputs by 83.5\%, 10.2\%, and 51.0\%, respectively.

{}

\begin{table}[]
\caption{Application throughputs of the users.}
\label{table:tp}
\begin{tabularx}{\linewidth}{l|l|L}
\hline
\textbf{MC setting} & \textbf{Mean throughput {[}kbps{]}} & \textbf{5th percentile throughput {[}kbps{]}} \\ \hline
MC Off              & 1360.5                          & 202.3                                     \\ 
RSRP-based           & 1452.5                          & 336.7                                     \\ 
BO-based          & 1534.0                          & 245.8                                     \\ 
MCS-based             & 1636.4                          & 371.2                                     \\ \hline
\end{tabularx}
\end{table}
\section{Conclusions}
\label{sec:conclusions}

In this paper, MC involving NTN was discussed. Further, a MC activation algorithm that aims to provide a throughput boost for the users in need was introduced. The developed algorithm was evaluated by system-level simulations in a scenario with a transparent payload LEO satellite and TN with nine gNBs. The results show that using MC with the introduced algorithm improves both the mean application throughput and 5th percentile throughput compared to when MC is turned off. Moreover, the introduced algorithm outperformed the baseline algorithm as well as a previously developed algorithm.

The introduced algorithm performs well for throughput enhancement. However, reliability aspects shall be considered in the future as well. Also, as the standardization progresses towards 6G, MC with regenerative payloads shall be considered. Yet another aspect to consider could be MC in multi-orbit satellite scenarios. Also, accounting for the different propagation delays in MC involving NTN should be considered.

\section*{Acknowledgment}
This work has been funded by the European Union Horizon-2020 Project DYNASAT (Dynamic Spectrum Sharing and Bandwidth-Efficient Techniques for High-Throughput MIMO Satellite Systems) under Grant Agreement 101004145. The views expressed are those of the authors and do not necessarily represent the project. The Commission is not liable for any use that may be made of any of the information contained therein.

\vspace{6pt}
\bibliography{references} 
\bibliographystyle{IEEEtran}

\end{document}